
\input phyzzx
\input epsf
\batchmode\font\eightcp=cmcsc8\errorstopmode
\ifx\eightcp\nullfont
	\font\eightcp=cmr8
\fi
\def\runningheads#1{\paperheadline={\iffrontpage\else
	{\hfil{\eightcp #1}\hfil}\fi}}
%
%

\def\refout{\par\penalty-400\vskip\chapterskip
	\spacecheck\referenceminspace
	\ifreferenceopen\Closeout\referencewrite\referenceopenfalse\fi
	\line{\fourteenrm\hfil REFERENCES\hfil}\vskip\headskip
	\offinterlineskip\tenpoint\input\jobname.refs}

\catcode`\@=11
\newfam\msbfam
\batchmode\font\twelvemsb=msbm10 scaled\magstep1 \errorstopmode
\ifx\twelvemsb\nullfont
	\def\Bbb{\bf}
	\message{msb fonts not found. 
		Redefining blackboard bold as boldface.}
\else	\font\tenmsb=msbm10 \font\sevenmsb=msbm7 
	\textfont\msbfam=\twelvemsb
	\scriptfont\msbfam=\tenmsb \scriptscriptfont\msbfam=\sevenmsb
	\def\Bbb{\relax\ifmmode\expandafter\Bbb@\else
	 	\expandafter\nonmatherr@\expandafter\Bbb\fi}
	\def\Bbb@#1{{\Bbb@@{#1}}}
	\def\Bbb@@#1{\fam\msbfam\relax#1}
\fi
	\catcode`\@=\active
\newfam\eusmfam
\newfam\scrfam 
\batchmode\font\tenscr=rsfs10 scaled \magstep1 \errorstopmode
\ifx\tenscr\nullfont 
	\catcode`\@=11
	\batchmode\font\twelveeusm=eusm10 scaled\magstep1 \errorstopmode
	\ifx\twelveeusm\nullfont
		\def\scr{\it}
		\message{rsfs fonts not found. Replaced with italic.}
	\else	\font\teneusm=eusm10 \font\seveneusm=eusm7 
		\textfont\eusmfam=\twelveeusm
		\scriptfont\eusmfam=\teneusm 
		\scriptscriptfont\eusmfam=\seveneusm
		\def\scr{\n@expand\f@m\eusmfam}
		\catcode`\@=\active
		\message{rsfs fonts not found. Replaced with eusm fonts.}
	\fi
\else	\font\sevenscr=rsfs7 scaled \magstep1 
	\font\fivescr=rsfs5 scaled \magstep1 
	\skewchar\tenscr='177 \skewchar\sevenscr='177 \skewchar\fivescr='177
	\textfont\scrfam=\tenscr \scriptfont\scrfam=\sevenscr
	\scriptscriptfont\scrfam=\fivescr
	\def\scr{\fam\scrfam}
\fi
\catcode`\@=11

\def\sss{\scriptscriptstyle}
\def\punkt{\quad .}
\def\komma{\quad ,}
\def\is{\!=\!}
\def\!{\mskip-.7\thinmuskip}
\def\l{\lambda}
\def\w{\omega}
\def\g{\gamma}
\def\Z{{\Bbb Z}}
\def\R{{\Bbb R}}
\def\C{{\Bbb C}}
\def\H{{\Bbb H}}
\def\O{{\Bbb O}}
\def\K{{\Bbb K}}
\def\a{\alpha}
\def\b{\beta}
\def\d{\delta}
\def\da{\d_\a}
\def\db{\d_\b}
\def\x{\xi}
\def\y{\eta}
\def\D{{\scr D}}
\def\tr{{\rm tr}}
\def\c{\!\circ\!}
\def\pro#1{\!\Buildrel 
	\raise 4pt\hbox{\the\scriptscriptfont0 #1}\under\circ\!}
\def\arrover#1{
	\vtop{	\baselineskip=8pt\lineskip=2pt
		\ialign{  ##\cr
			$\longrightarrow$ \cr
                	\hfill ${\scriptstyle #1}$\hfill \cr   }}}
\runningheads{M.~Cederwall, ``Introduction to Division Algebras,
	Sphere Algebras and Twistors''}
\date={October, 1993}
\pubnum={\vbox{\hbox{G\"oteborg-ITP-93-43}
		\hbox{hep-th/9310115}}}
\titlepage
	\vskip 1cm
	\epsfxsize=9cm
	\hbox{
		\hskip 2.9cm
		\epsffile{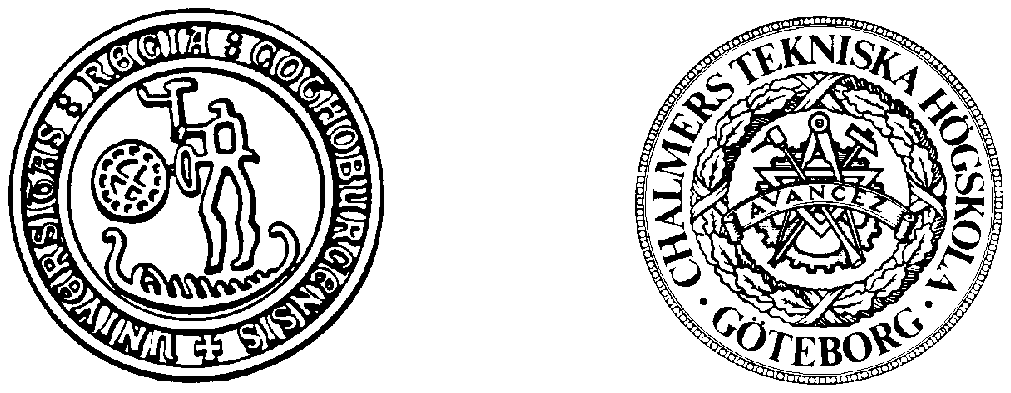}
		\hfill}
\vfill
\title{\break{\fourteenpoint Introduction to Division Algebras,
	Sphere Algebras and Twistors}}
\author{Martin Cederwall\foot{e-mail tfemc@fy.chalmers.se}}
\address{Institute for Theoretical Physics\break
	Chalmers University of Technology and University of G\"oteborg\break
	S-412 96 G\"oteborg, Sweden}
\vfill\abstract
A very basic introduction is given to the r\^oles of division algebras
and the related sphere algebras concerning the structure of space-time
in the dimensionalities $D\is 3,4,6$ and $10$, with special emphasis
on twistors transformations for light-likeness conditions and Hopf maps,
together with some outlook for particle and string theory.
\vfill
\centerline{\tenit Talk presented at the Theoretical Physics Network Meeting
	at NORDITA, Copenhagen, September 1993}
\vfill\endpage	
\REF\twistor{R.~Penrose and M.A.H.~McCallum
	\sl Phys.Rep. \bf 6 \rm (1972) 241
	\nextline\indent and references therein.}
\REF\Sudbery{A.~Sudbery, \sl J.Phys. \bf A17 \rm (1984) 939.}
\REF\Hurwitz{A.~Hurwitz, {\sl Nachr.Ges.Wiss.G\"ottingen} (1898) 309.}
\REF\BottMilnor{R.~Bott and J.~Milnor, {\sl Bull.Am.Math.Soc.} 
	{\bf 64} (1958) 87.}
\REF\Schafer{R.D. Schafer, An introduction to
	nonassociative algebras, New York (1964).}
\REF\Cayley{A.~Cayley, \sl Phil.Mag. \bf 3 \rm (1845) 210.}
\REF\htwistor{I.~Bengtsson and M.~Cederwall, \sl Nucl.Phys. 
	\bf B302 \rm (1988) 81.}
\REF\tentwistor{N.~Berkovits, \sl Phys.Lett. \bf 247B \rm (1990) 45;\nextline
	M.~Cederwall \sl J.Math.Phys. \bf 33 \rm (1992) 388.}
\REF\JordanJordan{P.~Jordan, \sl Nachr.Ges.Wiss.G\"ottingen \rm (1932) 569.}
\REF\Cartan{E.~Cartan and J.A.~Schouten, {\sl Proc.Kon.Wet.Amsterdam}
	{\bf 29} (1926) 803; 933.}
\REF\Wolf{J.A.~Wolf, {\sl J.Diff.Geom.} {\bf 6} (1972) 317; 
	{\bf 7} (1972) 19.}
\REF\ESTPS{F. Englert, A. Sevrin, W. Troost, A. Van
Proyen and Ph. Spindel,\nextline\indent {\sl J. Math. Phys} {\bf 29} (1988)
281.}
\REF\sevensphere{M.~Cederwall and C.R.~Preitschopf,
	G\"oteborg-ITP-93-34, hep-th/9309030}
\REF\Hopf{H.~Hopf, \sl Math.Ann. \bf 104 \rm (1931) 637,\nextline\indent
	 \sl Fund.Math. \bf XXV \rm (1935) 427.}
\REF\PenRind{R.~Penrose and W.~Rindler, ``Spinors and Space-time,
	volume 2'', Cambridge University Press (1986).}
\REF\Berkovitsstring{N. Berkovits, \sl Nucl.Phys.\bf B358 \rm (1991) 169.}
\REF\galsok{
	A. Galperin and E. Sokatchev,
	\sl Phys.Rev\bf D46 \rm (1992) 714;
\nextline\noindent F. Delduc, E. Ivanov and E. Sokatchev, 
	\sl Nucl.Phys.\bf B384 \rm (1992) 334;
\nextline\noindent F. Delduc, A. Galperin, P. Howe and E. Sokatchev, 
	\nextline\indent\sl Phys.Rev.\bf D47 \rm (1993) 578.}
\REF\string{M. Cederwall, \sl Phys.Lett.\bf 226B \rm (1989) 45.}
\REF\Eightconf{L.~Brink, M.~Cederwall and C.R.~Preitschopf,
	\sl Phys.Lett.\bf B311 \rm (1993) 76.}
\REF\jordanmech{M.~Cederwall, \sl Phys.Lett. \bf 210B \rm (1988) 169.}
This introduction, given the condition that it should fit into
a single lecture, will be quite schematical. Many mathematical
and technical details will be left out, for the purpose of giving
a more intuitive overview of the subject. The intention is that the
listener/reader should get a starting point for more detailed or
extensive study. There will be nothing said
about supersymmetry, although its incorporation in the context of
twistors is very natural and interesting.

The basic object to start with is a 
null vector $P^\mu$ in Minkowski space. It can be thought of as the
momentum for a massless particle. It fulfills
$$P^2=0\punkt$$
In four dimensions, $P^\mu$ may be expressed as a bilinear of spinors:
$$P^\mu={1\over 2}\l\g^\mu\l\punkt\eqn\twist$$
This is a ``twistor transformation'' [\twistor].
This is most easily seen using the isomorphism $SO(1,3)\approx SL(2;\C)$
[\twistor,\Sudbery]. 
Then, a spinor
is a two-component complex object
$$\l=\left[\matrix{\x\cr\y\cr}\right]\komma$$
and a vector is a hermitean 2$\times$2-matrix
$$P=\left[\matrix{\sqrt 2p^+&p^*\cr p&\sqrt 2p^-\cr}\right]
	=\,\hbox{''}{P_\mu\g^\mu}\, \hbox{''}\punkt$$
The gamma matrices are
$$\g^+=\left[\matrix{\sqrt 2&0\cr 0&0\cr}\right]\komma\quad
	\g^-=\left[\matrix{0&0\cr 0&\sqrt 2\cr}\right]\komma\quad
	\g^I=\left[\matrix{0&e^*_I\cr e_I&0\cr}\right]\punkt\eqn\gammam$$
where $e_I$, $I\is 0,1$, forms a basis for the complex numbers.
The spinor bilinear in \twist
$$P=\l\l^\dagger=\left[\matrix{\x\x^*&\x\y^*\cr\y\x^*&\y\y^*}\right]$$
is easily seen to be light-like. In $SL(2;\C)$ language, this reads
$$P^2=P\tr P\punkt$$

For which dimensionalities is it possible to generalize these
statements?
If we take the gamma matrices formally as in \gammam, where $\{e_I\}$
is a basis for the transverse space, we obtain two conditions:

\item{\bullet} Existence of Clifford algebra $\,\Longrightarrow\,$ 
	Alternative algebra,
\item{\bullet} Light-likeness of spinor bilinear $\,\Longrightarrow\,$ 
	Division algebra.

To clarify, a (normed) division algebra [\Schafer] is an algebra 
(not necessarily associative)
where $|ab|=|a||b|$, and alternativity means that $[a,b,c]\equiv
(ab)c-a(bc)$ is completely antisymmetric.
There is a theorem [\Hurwitz,\BottMilnor,\Schafer], 
that the only (real) alternative division algebras are
$$\eqalign{&\R\komma\quad\hbox{the real numbers},\cr
&\C\komma\quad\hbox{the complex numbers},\cr
&\H\komma\quad\hbox{the quaternions},\cr
\hbox{and}\quad&\O\komma\quad\hbox{the octonions [\Cayley]}.\cr}$$
I denote them $\K_\nu$, where $\nu=1,2,4,8$ are the dimensionalities.
The uniqueness of these algebras is the fundamental reason for the 
well known statement
$$(\l_1\g_\mu\l_2)\g^\mu\l_3+\hbox{cyclic permutations}=0$$
in $D=\nu+2=3,4,6,10$,
which is a stronger version of the light-likeness of the spinor bilinear.
These dimensionalities are the ones where twistor transforms exist
[\twistor,\htwistor,\tentwistor]. 

Here follows some notation and properties of the division algebras:

\noindent
Orthonormal basis: $\K_\nu\ni x=\sum_{I=0}^{\nu-1}x_Ie_I\quad(e_0=1)\,.$
	\nextline
Conjugation ($x\rightarrow x^*$): $e_0\rightarrow e_0$, 
	$e_i\rightarrow -e_i\,,i=1,\ldots,\nu-1\,.$
	\nextline
Norm: $|x|=(x^*x)^{1/2}\,.$\nextline
Real part: $[x]={1\over 2}(x+x^*)\,.$\nextline
Imaginary part: $\{x\}={1\over 2}(x-x^*)\,.$\nextline

\noindent
Multiplication table (illustrated for $\O$ in the figure below, where moving 
clockwise in the triangle gives positive sign, and the triangle may
be rotated any integer multiple of ${2\pi\over 7}$):
$$\eqalign{&e_ie_j=-\delta_{ij}+\sigma_{ijk}e_k\,,		\cr
	&\sigma\hbox{ antisymmetric, }\sigma_{i,i+1,i+n}=1\hbox{ where }
		\nu=2^n\,,					\cr
	&e_{i+\nu-1}=e_i\,.\cr}$$
\epsfysize=.4\vsize\hskip2cm
\vbox to .4\vsize{\epsffile{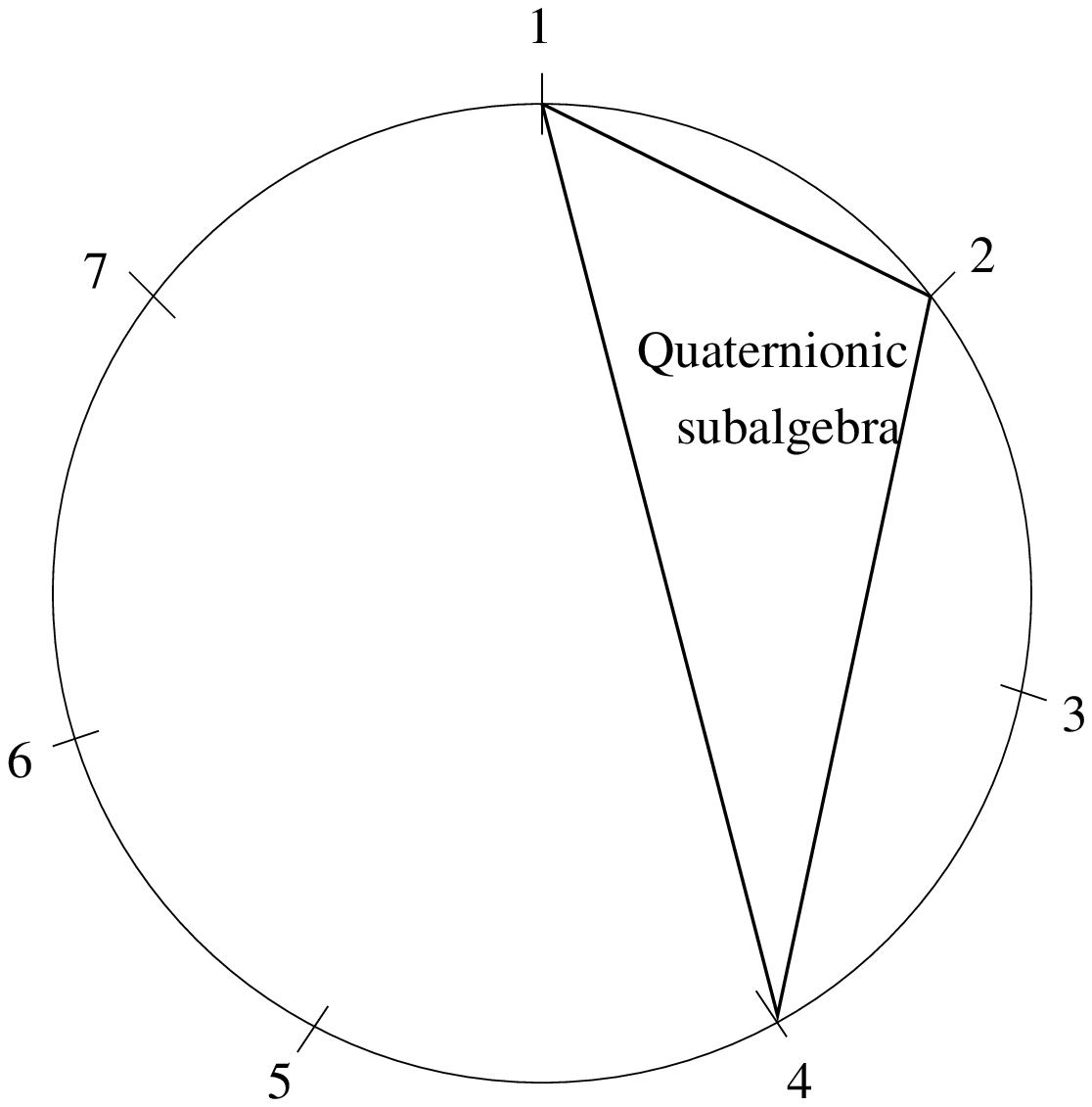}}\nextline\hskip-1cm
$\R$, $\C$ commutative, associative.\nextline
$\H$ non-commutative, associative.\nextline
$\O$ non-commutative, non-associative (but alternative);\nextline

\noindent
$[e_i,e_j,e_k]=2\rho_{ijkl}e_l$, $\rho_{ijkl}=-(^*\sigma)_{ijkl}=
		-{1\over 6}\epsilon_{ijklmnp}\sigma_{mnp}$.

\vskip.5cm\underbar{Space-time and division algebras:}

The isomorphism $SL(2;\C)\approx SO(1,3)$ is generalized to [\Sudbery]
$$SL(2;\K_\nu)\approx SO(1,\nu+1)\punkt$$
Also the conformal algebras of $(\nu+2)$-dimensional space-time
are naturally formulated as 
$$Sp(4;\K_\nu)\approx SO(2,\nu+2)\punkt$$

There is a natural interpretation in terms of 
Jordan algebras [\JordanJordan]:
A Jordan algebra has a symmetric product fulfilling
$$X\c((X\c X)\c Y)=(X\c X)(X\c Y)\punkt$$ 
The $n\!\times\!n$ hermitean
matrices with division algebra entries is a basis for a Jordan
algebra for all $n$, $\nu\neq 8$, and for $n\leq 3$, $\nu=8$.
This means that the vectors 
$$V=\left[\matrix{\sqrt 2v^+&v^*\cr v&\sqrt 2v^-\cr}\right]$$
in $(\nu+2)$-dimensional Minkowski
space with product $U\c V={1\over 2}(UV+VU)$ form a Jordan algebra. 
The Lorentz group is its structure group,
and the conformal group its conformal group.

\vskip.5cm\underbar{Light-like lines and projective spaces:}

The celestial sphere in $\nu+2$ dimensions (the space of light-cone
directions) is $S^\nu$.

\vskip-.5cm
\epsfxsize=.4\vsize 
\hskip1cm\vbox to .4\vsize{\epsffile{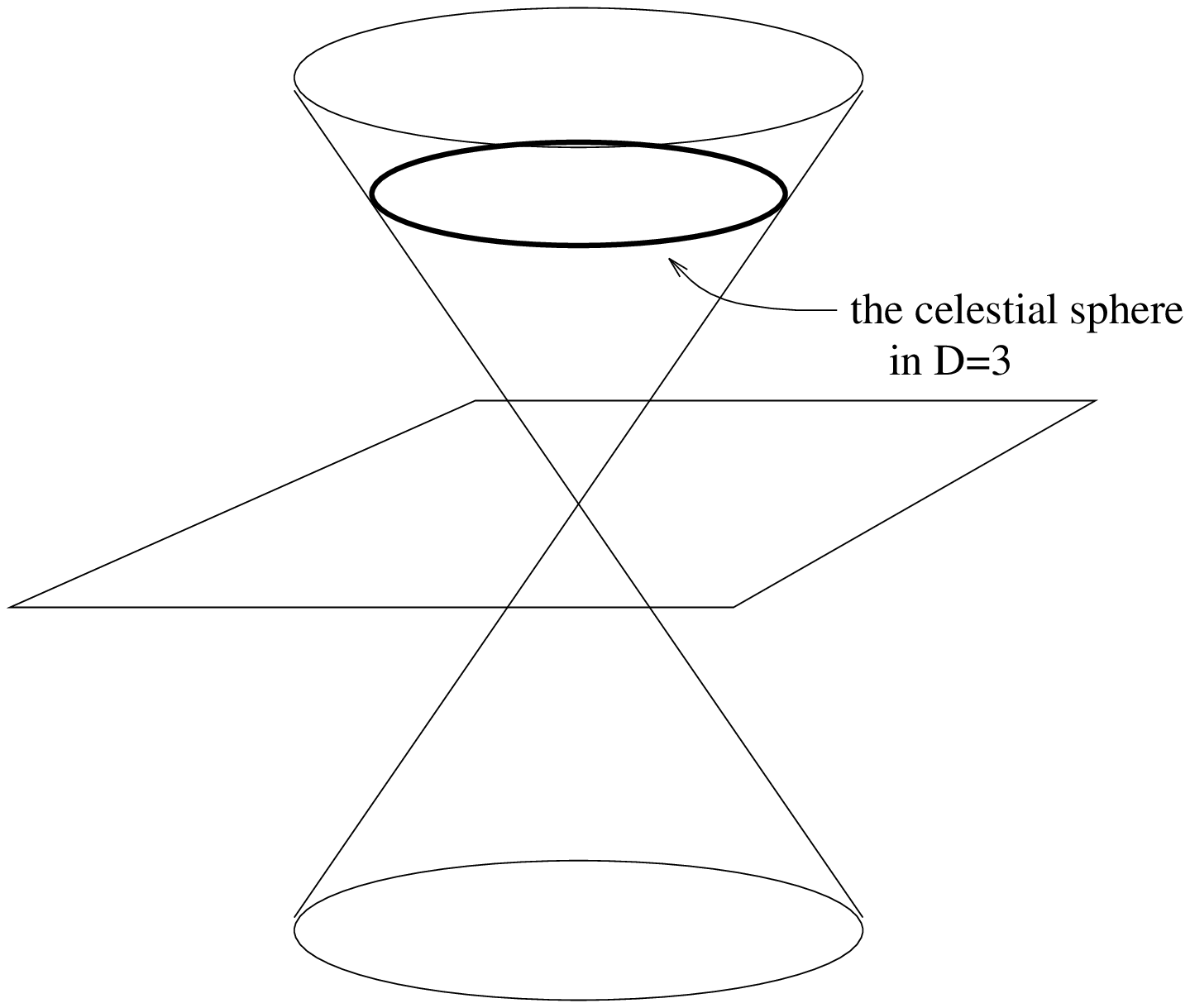}}\vskip-.5cm

We have two ways of representing it

\item{\bullet} $\{P\ |\ P^2=P\tr P\}\,,$
\item{\bullet} $\{\l\ \hbox{modulo transformation leaving }\l\l^\dagger
				\hbox{ invariant}\}\,.$ 

The first one is a well known representation of the projective space
$\K_\nu P^1=S^\nu$. To understand the second one, we must examine the
invariance of $\l\l^\dagger$.
$$\eqalign{\delta(\x\x^*)=0&\ \Longrightarrow\ \x\rightarrow\x\Omega\,,\quad
				|\Omega|=1\,,\cr
	\delta(\y\x^*)=0&\ \Longrightarrow\ \y\rightarrow
		\y\pro X\Omega\,,\cr}\eqn\sphere$$
where $a\pro Xb=(a(bX^*))X=(aX^*)(Xb)=X^*((Xa)b)$. 

The set of transformations (parametrized by $\Omega$) obviously
has the topology $S^{\nu-1}$. For the cases $\nu=1,2,4$ the groups
are $S^0=\Z_2$, $S^1=U(1)$ and $S^3=SO(3)$, realized by 
(right) multiplication
by unit elements in $\K_\nu$. For the case $\nu=8$, we have the
space $S^7$, which, due to non-associativity, is not a group.

However, the transformations \sphere\ close to a ``group''!
With the infinitesimal version $\da\x=\x\a$, $\da\y=\y\pro X\a$ ($[\a]=0$):
$$[\da,\db]=\d_{[\a,\b]_X}\komma\eqn\algebra$$
where $[a,b]_{\sss X}=a\pro Xb-b\pro Xa$.

Explanation: There is a classification of ``parallelizable'' 
manifolds [\Cartan,\Wolf].
Parallelizable here means that there are as many globally defined
orthonormal vectorfields as the dimension of the manifold.

\item{}Examples: $S^1$ is a trivial example, where the vector field is 
	tangent to the circle. $S^2$, like any even-dimensional sphere, 
	does not have any vectorfield on it -- it can not be ``combed''.

The only simply connected compact parallelizable manifolds are the
Lie groups and $S^7$. If these vectorfields exist one can use them
to define parallel transport of vectors. Since transport around any
closed curve gives back the same vector, the curvature of the corresponding
connection vanishes. We can think of the manifold equipped with this 
connection as ``flat'', and the transport as translation.

If the parallelizing connection is written $\tilde\Gamma=\Gamma-T$,
where $\Gamma$ is the metric connection, the vielbeins will not be
covariantly constant, but transport as $\D e=T$ ($T$ is {\it torsion},
and this can be taken as its definition). Then it follows
from $[\D_m,\D_n]=R_{mn}$ that 
$$[\D_a,\D_b]=2{T_{ab}}^c\D_c$$
($m,n$ space indices, $a,b$ tangent indices).

These are our $S^7$ transformations [\ESTPS,\sevensphere]. 
What distinguishes $S^7$ from the
Lie groups is that its torsion (``structure constants'') vary over
the space. Explicitely, one choice of the torsion tensor is the one
in \algebra:
$$T_{ijk}(X)=[e^*_i(e_j\pro Xe_k)]\,.$$

So, back to the spinor realization of the celestial sphere $S^\nu$!
We found that $\l\l^\dagger$ is invariant under $S^{\nu-1}$ tranformations.
The celestial sphere is the space of $S^{\nu-1}$ orbits in $\R^{2\nu}$
modulo a positive real scale, \ie\ in $S^{2\nu-1}$. The modding out
of $S^{\nu-1}$ is the Hopf map [\Hopf].
$$\eqalign{S^{1} \ & \arrover {S_0} \ S^1 = \R P^1 \cr
	S^{3} \ & \arrover {S^1} \ S^2 = \C P^1 \cr
	S^{7} \ & \arrover {S^3} \ S^4 = \H P^1 \cr
	S^{15}\ & \arrover {S^7} \ S^8 = \O P^1 \ . \cr}
		\eqn\hopfmaps$$

Equivalently, the total space $S^{2\nu-1}$ is a (topologically non-trivial)
fiber bundle over $S^\nu$ with $S^{\nu-1}$ as fiber.

{\it The twistor transform is the translation between the Jordan algebra 
and homogeneous coordinates descriptions of} $\K_\nu P^1$.

When formulated in terms of orbits in a two-component $\K_\nu$-valued object
(the spinor) the maps look simple. The topology is quite non-trivial -- we
take the complex Hopf map as example. Any $S^1$ orbit is linked to
any other. We refer to reference [\PenRind] for an illustration.

\vskip.5cm\underbar{Particles:}

One passes to twistor variables with the transform
$$\matrix{&P=\l\l^\dagger	&(\ P^\mu={1\over 2}\l\g^\mu\l\ )\cr
	&\w=X\l			&(\ \w=X_\mu(\g^\mu\l)\ )\cr}$$
The spinor $\w$ is conjugate to $\l$.
Counting the degrees of freedom:
$$\eqalign{(X,P)\ &:\ (\nu+2)-2\cdot 1=2(\nu+1)\cr
	(\l,\w)\ &:\ 2\cdot 2\nu-2(\nu-1)=2(\nu+1)\cr}$$

The twistor variables form together 
$$Z=\left[\matrix{\l\cr\w\cr}\right]$$
which is a spinor of the conformal group, $Sp(4;\K_\nu)$. The conformal
symmetry can be made manifest (it can in a space-time picture, too).
From $Z$, $S^{\nu-1}$ generators can be formed.

\vskip.5cm\underbar{Strings:}

The same equations hold. If we want to treat left- and rightmovers
separately, $P$ and $X$ are not independent -- $P=\partial X$. 
This gives rise to
constraints between $\l$ and $\w$ (spoiling conformal invariance). 
These are much like
the spinorial constraints for the superparticle or superstring:
half first class, half second class [\Berkovitsstring]. 
The first class ones are the
Virasoro constraint and an $S^{\nu-1}$ Kac-Moody constraint.\nextline

Counting degrees of freedom:
$$\eqalign{X\ &:\ (\nu+2)-2\cdot 1=\nu\cr
	(\l,\w)\ &:\ 2\cdot 2\nu-1\cdot {1\over 2}\cdot 2\nu-
		2\cdot 1-2\cdot(\nu-1)=\nu\cr}$$
The second class constraints pose problems for covariant quantization.
The twistor formulation, that solves this problem for the superparticle,
has just the same obstacle to covariant quantization in the
bosonic sector! 

There are different $D=10$ twistor-like approaches
[\galsok\ and references therein],
where one takes a set of eight twistors instead of one.
Then the KM symmetry is $\widehat{SO(8)}$ instead of $\widehat{S^7}$.
I don't think anyone has performed a canonical analysis on these,
but the problem with second class constraints remain.

There is an alternative string twistor formulation [\string], where left- and
rightmovers mix (off shell), and where all constraints are first class.
It is reached by using both $P$ and $X$,
and replacing $P$ by the sum $\l_1\l_1^\dagger+\l_2\l_2^\dagger$.
Then, nothing can be said about $P$ except that it is light-like
or time-like. Apart from the two $S^{\nu-1}$'s there is a $\nu$-
dimensional mixture between the two $\l$'s. The manifolds of $\l$'s
giving the same $P$ can be shown to be $U(1)$, $SO(3)\times U(1)$,
$SO(5)$ and $SO(9)/G_2$. The last one is not parallelizable, so
it is unclear if this can be used in $D=10$.

Even though the problems with the second class constraints remain,
the twistor formulation of superstrings has an interesting structure.
The Kac-Moody algebra $\widehat{S^{\nu-1}}$ is enlarged to an 
$N=\nu$ superconformal algebra. There is a theorem that SCA's do not
exist for $N>4$, but this is circumvented by the field-dependence of
the $S^7$ structure functions (the torsion tensor).

Generators:
$$\matrix{1\ &\hbox{spin 2}\cr
	\nu\ &\hbox{spin 3/2}\cr
	\nu-1\ &\hbox{spin 1}\cr}$$

An interesting thing is that these algebras are present also in the
light-cone superstring [\Eightconf], that seems to have nothing to do with twistors!
One may hope, that it survives as a remnant of a gauge fixed symmetry,
like the Virasoro algebra in the light-cone bosonic string.

The algebra $\widehat{S^7}$ [\sevensphere] has some peculiar features that
are not pre\-sent in the lower-dimensional algebras. The anomalies may
be field-dependent, but a redefinition of the current yields a central
extension of the ordinary type, with one certain numerical coefficient
disregarding the field content!
For any field content, the BRST operator can be made to vanish
quantum mechanically. We do not yet understand at what point the
algebra can be used to make predictive statements about field content.
We also lack a superfield formulation for the $N=8$ superconformal algebra.
 
The Jordan algebra business can be carried on to $3\times3$ matrices
of octonions, and to matrices of any dimension for the other 
division algebras [\Sudbery,\jordanmech].
The ``exceptional'' Jordan algebra $J_3(\O)$ is of exceptional interest, 
since it has exceptional symmetry groups.\nextline

Analogies:\vskip.3cm

\hskip1cm	
\vbox to 3cm{\tabskip=0pt\offinterlineskip\def\t{\noalign{\hrule}}
	\halign{\strut#\hfil&\quad\vrule\quad#&\hfil#\hfil&\quad\vrule\quad#
		&\hfil#\hfil\cr
	&&$J_2(\O)$&&$J_3(\O)$\cr\t
	derivations:&&$SO(9)$&&$F_4$\cr\t
	structure:&&$SO(1,9)$&&$E_6$\cr\t
	conformal:&&$SO(2,10)$&&$E_7$\cr}}

The matrices $P\in J_3(\O)$ with $P^2=P\tr P$ modulo a real positive
scale form the projective space $\O P^2$. The lack of homogeneous
coordinates for $\O P^2$ seems to make a twistor transform impossible.
We have however constructed objects in $\O^3$ that seem to be
natural generalizations of homogeneous coordinates [\sevensphere]. 
They {\it may}
be useful for constructing a twistor transform for $J_3(\O)$, and
{\it may} have transformation properties under $E_6$ that are
``almost spinorial''. {\it Maybe} (non-linear) super-extensions of
exceptional groups with $S^7$ as the remaining bosonic generators
can be found, and {\it maybe} also a superconformal algebra in $D=10$
can be formulated this way... 

\refout
  
\end